\documentclass[prl,
a4paper,
showpacs, 
twocolumn
]{revtex4}
\usepackage{graphicx}
\def\eq#1{{Eq.~(\ref{#1})}}

\newcommand{\be}{\begin{equation}}
\newcommand{\ee}{\end{equation}}
\newcommand{\bea}{\begin{eqnarray}}
\newcommand{\eea}{\end{eqnarray}}

\begin{document}
\title[Non-Newtonian gravity]
{Non-Newtonian gravity confronting models with large extra dimensions}
\author{M.~Fabbrichesi}
\author{M.~Piai}
\author{G.~Tasinato}
\affiliation{INFN, Sezione di Trieste and\\
Scuola Internazionale Superiore di Studi Avanzati\\
via Beirut 4, I-34014 Trieste, Italy.\\}
\begin{abstract}
We discuss the interplay between direct constraints on non-Newtonian
gravity and  particle-physics
bounds in models with large extra dimensions. 
Existing and future bounds and the most effective ways of further testing these
models in gravitational experiments are compared and discussed. 
\end{abstract}
\maketitle

Additional space dimensions 
large enough to be observable have been
suggested~\cite{extra,add} as a possible solution to the problem of
the large scale difference between gravity and the standard model. 
In this approach the standard-model particles are bound within 
four-dimensional space time~\cite{RS} and---along the lines of
brane-world models~\cite{HW}---only gravity inhabits the extra
dimensions $\delta$. 

Gravity is 
 modified at short distances and becomes much stronger; 
its experimental long-distance
weakness is  explained  by the
space volume of the large extra dimensions by which
the coupling must be divided in four dimensions.
 The  length scale $R^*$ at which the long distance regime is 
recovered  is model dependent, and all possibilities between the
millimeter and the inverse Planck mass are in principle available.

Such a scenario has spurred new interest
in experimental tests of the short-distance 
behavior of gravity in the hope of either finding deviations 
or at least setting bounds on the fundamental
 coupling of gravity at lengths below the millimeter.

In this letter we discuss the complementary role of two classes
 of such experimental tests: particle-physics measurements---sensitive
to the strength of the coupling independently of distance---and 
short-distance measurements of Newton law---sensitive 
to the range of non-Newtonian effects down to nanometers. 
In the discussion, the 
different role played by bounds derived assuming a specific
 compactification geometry and those that are compactification
 independent is stressed. 

Our conclusion is that particle-physics measurements
already rule out  for  $\delta > 2$ 
the possibility that non-Newtonian effects
induced by models
with large extra dimension can be found in gravimetric experiments
below $10^{-5}$ m.
This result is independent of the compactification model used. 
Also the cases
$\delta =1$ and 2 are excluded  if
compactification on a $\delta$-torus is chosen. The best  window for
confronting non-Newtonian gravity and models with large extra
dimensions is therefore at, and just below, the mm range.

Even though cosmological and astrophysical constraints can also be
important, and  very 
strong for the case of two large extra dimensions~\cite{astro}, 
we will not discuss them here. Similarly, we
do not discuss models in which the
standard-model particles are also 
allowed to propagate into the extra dimensions~\cite{Anton}
or models with non-factorizable geometries~\cite{Lisa}.
\vskip1.5em
{\bf Searches for non-Newtonian gravity.}
While Newtonian gravity above the centimeter is well confirmed~\cite{AdR},
its short distance behavior is still under active scrutiny.
All experiments, regardless of the actual apparatus, 
set a bound on non-Newtonian interactions from the absence of
deviations between the force measured at distance $r^{\ast}$ and
the predicted one. 

The bound is usually given in terms of the parameters $\alpha$ and $\lambda$
according to the two-body potential
\be
\left. V(r)\right|_{r^*}= \left. \frac{G_N m_1 m_2}{r} \Biggl[ 
1+\alpha_{G}\, e^{-r/\lambda}  \label{pot_exp}
\Biggr]\right|_{r^*} \, ,
\ee
where $G_N$ is the Newton constant in four space-time dimensions.
Because of  the exponential behavior,
the best sensitivity is achieved in the range $\lambda \sim r^{\ast}$.
Currently,
experiments testing Van der Waals forces  are sensitive to the range 
$r^{\ast} \sim 1.5 \div 130$ nm~\cite{vdW};
Casimir-force experiments explore 
$r^{\ast} \sim 0.02 \div 6$ $\mu$m , $r^{\ast}$ being here the
distance between dielectrics or metal
surfaces~\cite{casimir1,casimir2} or  up  to mm
by means of a torsion pendulum\cite{casimir3}.
Cavendish-type experiments, in which the gravitation force is directly
measured, are sensitive to $r^{\ast} > 1$ mm~\cite{torsion}. 

\begin{figure}[ht]
\includegraphics[scale=0.5]{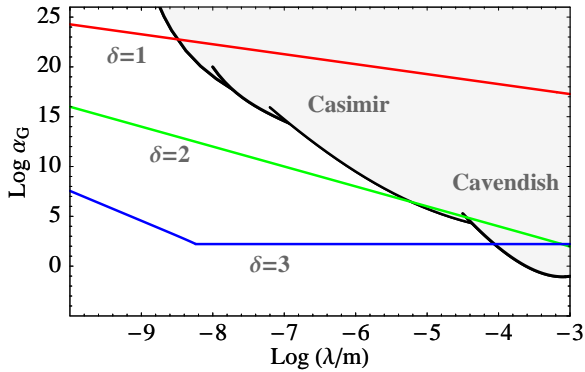}
\caption{Bounds $\alpha_G$ vs.\ $\lambda$ for $\delta =$ 1, 2 and 3 from
current particle-physics tests (see Table~\ref{hep-bounds}). 
The thicker curved lines are the best available bound from
non-Newtonian gravity experiments and the grey area is the range
excluded by them. We take $R^* \simeq 10^{-3}$ m.}
\label{fig1}
\end{figure}

\begin{figure}[ht]
\includegraphics[scale=0.5]{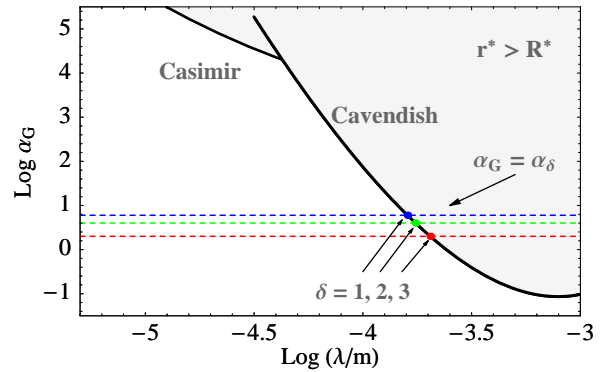}
\caption{LD bounds $\alpha_G$ vs.\ $\lambda$. Dots represent the
intersection between $\alpha_G=\alpha_\delta$ and the experimental
bounds.
All extra dimensions are compactified on a $\delta$-torus, which gives
$\alpha_\delta = 2\, \delta$~\cite{expo}.
The resulting bounds  are 
 $M_f \agt 5.3 \times 10^{5}$ TeV for $\delta = 1$, 
$M_f \agt 3.7$ TeV and $M_f \agt 3.1$ GeV, respectively, 
in  $\delta = 2$ and 3. }
\label{fig2}
\end{figure}

\begin{figure}[ht]
\includegraphics[scale=0.5]{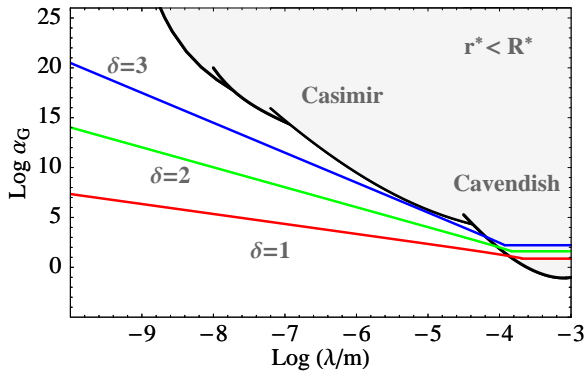}
\caption{
SD bounds $\alpha_G$ vs.\ $\lambda$. 
The curves are computed at  $M_f \agt 5.3 \times 10^{5}$ TeV for $\delta = 1$, 
$M_f \agt 3.7$ TeV and $M_f \agt 3.1$ GeV, respectively, 
in  $\delta = 2$ and 3. The change of slope takes place at $\lambda_{max}$.}
\label{fig3}
\end{figure}

\begin{figure}[ht]
\includegraphics[scale=0.5]{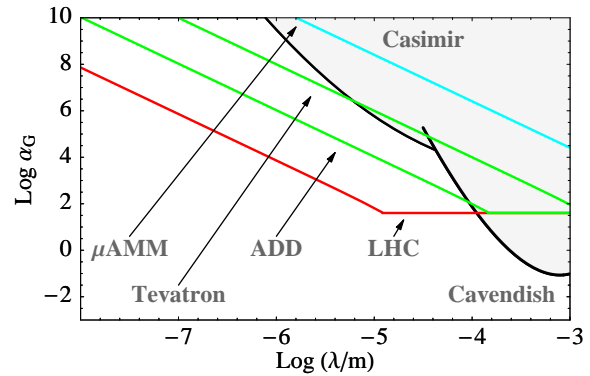}
\caption{Bounds $\alpha_G$ vs.\ $\lambda$ for $\delta =2$: comparison between 
the sensitivity of all available techniques, including planned LHC reach.}
\label{fig4}
\end{figure}

The exclusion regions
thus determined are convex curves around the 
distance $r^*$ at which the experiment is
performed, rapidly 
 becoming less sensitive at smaller separations.
The combined exclusion regions obtained by these searches, for the relevant
distances, are shown as grey areas delimited by black curves in 
Figs.~\ref{fig1}-\ref{fig4}. 
\vskip1.8em
{\bf Gravitational potential in models with large extra dimensions.}
The two-body potential in models with extra dimensions is
parameterized (for $r$ less than $R^*$, 
the characteristic compactification length) as~\cite{add}
\be
V_\delta (r)=\frac{G_N m_1 m_2}{r} 
\left( \frac{a_\delta}{r}
\right)^{\delta} \label{pot_s} \, .
\ee 
In \eq{pot_s} 
\be
a_\delta =
(G^{(\delta)}/G_N)^{1/\delta}= \frac{2\,\pi}{M_{f}}\;
\left(\frac{4\,\pi}{\Omega_{\delta}}
\frac{M_{\rm P}^2}{M_{f}^{2}} \right)^{1/ \delta} \, , \label{adelta}
\ee
where we define
$M_{\rm P} \equiv 1/\sqrt{G_N} = 1.22 \times 10^{16}$ TeV. In \eq{adelta},
$\Omega_{\delta}=2 \pi^{(3+\delta)/2}/ \Gamma[(3+\delta)/2]$
and $M_f$ is the scale of the
effective theory. 
For distances larger than $R^*$,  
the potential in \eq{pot_s} is replaced by the usual
Newtonian potential plus exponentially small corrections:
\be
V_\delta (r) = 
\frac{G_N m_1 m_2}{r} \Biggl[ 
1+\alpha_\delta\, e^{-r/R^*} +  \cdots  \label{pot_l}
\Biggr]  \, .
\ee
In \eq{pot_l}, 
the value of $\alpha_\delta$ depends on the
compactification choice and is of the order of the number of 
extra dimensions~\cite{expo}. 

It is important to bear in mind that
\eq{pot_l} depends on the way the extra dimensions are
treated in the process of compactification
while \eq{pot_s} only relies on Gauss law and is therefore 
compactification independent. 

When the experimental bounds parameterized by \eq{pot_exp} 
are plotted (on a logarithmic scale in the
$(\alpha-\lambda)$-plane)
against \eq{pot_l}, a single point is obtained 
at $\alpha_G = \alpha_\delta$ and
$\lambda = R^*$; on the other hand, when the same bounds are compared with
\eq{pot_s}, they give lines,
 the shape and position of which are controlled by $M_f$ and the
number of large extra dimensions.

\vskip1.5em
{\bf Compactification-independent  bounds from particle physics.}
Contrarily to short-distance gravity measurements,
particle-physics measurements are independent of $r$ and only 
constrain the effective gravitational coupling $G^{(\delta)}$
by means of the bound on $M_f$.  

This independence from $r$ is manifest in the $(4+\delta)$-dimensional
 theory, which probes distances
much smaller than the compactification radius $R^*$, and recovered in the
 4-dimensional computation  after resumming over  the Kaluza-Klein states.

For this reason, the relationship
obtained by comparing \eq{pot_exp} and \eq{pot_s},
must be valid for any choice of $r$ (as long as $r \alt R^*$)
and gives the stringiest bound at the minimum.
 Therefore, the curve of exclusion is
found to be
\be
 \alpha_G (\lambda) \le \mathop{\mathrm{min}}_{r} \Big\{ \Big[ \left(\frac{a_{\delta}}{r} \right)^{\delta}-1 \Big]e^{r/\lambda} \Big\} \, .
 \label{5}
\ee
To find the curve given by \eq{5}, we must solve the polynomial
equations 
obtained by the minimalization procedure. 
Exact solutions exist for $\delta < 4$; however,
 for all practical purposes, approximated solutions can be found by elementary
calculus for any $\delta$. The exclusion region is given by the lines
\be
\alpha_{G}(\lambda) = \left\{ \begin{array}{ll}
\left[ \left( a_{\delta}/\delta \lambda \right)^\delta - 1 \right] e^\delta & 
\textrm{for $\lambda < \lambda_{max}$} \\ & \\
 \alpha_{min} \equiv \delta \, e^{\delta+1} &   
\textrm{for $\lambda \geq \lambda_{max}$} \, , 
\end{array} \right. \label{6}
\ee
where $\lambda_{max} = (1+\delta)^{-(1+\delta)/\delta}a_{\delta}$. The
value $\lambda_{max}$ is reached when no real solution can be
found. The exclusion region is extended for $\lambda > \lambda_{max}$ 
by taking   smaller values of $M_f$ (already excluded)  for which the solution is
translated to larger values of $\lambda$ while still ending at the same
(constant) value of $\alpha_{min}$.  

Recent calculations have considered precision measurements
like the anomalous magnetic moment of
the muon~\cite{mmm} and radiative 
oblique parameters~\cite{oblique}, as well as 
collider physics~\cite{colliders}. In collider physics,
the most effective channel at both LEP and Tevatron is that in which
virtual gravitons take part in dilepton or diphoton production. 
Production of real graviton gives less stringent bounds. Whenever the
bound depends on the sign of the potential we have taken the lesser bound.
Recent reviews of all these
bounds can be found in Ref.~\cite{rev}. 

We have summarized in
Table~\ref{hep-bounds} the best bounds from particle physics.
\begin{table}[ht]
\caption{Particle physics bounds on $M_f$. The numbers reported are
the constrains in TeV for the first few large extra dimensions
$\delta$. Missing entries were not reported in the literature.}
\label{hep-bounds}
\begin{tabular}{c|cccc|c}
\hline
 & \multicolumn{4}{c|}{$\delta$} &  \\
                   \cline{2-5}
measurement             & 1  & 2   & 3   & 4   &   reference               \\  
                    \cline{2-5}          
\hline
$(g_\mu-2)/2$&$\quad 0.3\quad$&$\quad 0.3\quad$&$\quad 0.4\quad$&$\quad 0.4\quad$&\cite{mmm}\\
\hline
LEP          &1.2& 1.2 & 1.2 & 1.2&  \cite{l3} \\
Tevatron I   &- &1.5  & 1.5 & 1.3&  \cite{d0,tevatron} \\
\hline
Tevatron IIb &- &3.5  & 3.0 & 2.6&  \cite{tevatron} \\
LHC          &- & 13 & 12 & 10&  \cite{tevatron} \\
\hline
\end{tabular}
\end{table}
While precision measurements and
collider bounds from production of real gravitons depend on the
number of extra dimensions, those from virtual graviton processes at
colliders are (almost,
for certain parameterizations) independent. Bounds from oblique parameters are
potentially very restrictive but are plagued by infrared divergences
which make the final result rather uncertain. For this reason we will
not use them. 

Even though a degree
of uncertainty remains in these calculations because of the cut-off 
dependence (and because of different parameterizations), 
the bounds work on order of magnitudes and are therefore sufficiently
reliable as they stand. In particular, we neglect small
discrepancies between different approaches~\cite{colliders}.

We keep in Table~\ref{hep-bounds} also the $\delta=$1
 case, even though it is often considered ruled out.
This is true only
after having assumed a specific compactification geometry 
and we want 
to use the particle-physics constraints irrespectively 
of this additional assumption.
\vskip1.5em
{\bf Non-Newtonian gravity and particle-physics constraints.}
Given the particle-physics bounds in Table~\ref{hep-bounds} and \eq{6},
we obtain the curves in Fig.~\ref{fig1}, where
the respective exclusion regions (the
area above the lines) are presented for the first few extra dimensions. 
In using these bounds, 
the values for $\alpha_G$ and
$\lambda$ of a specific model
must be plotted against the bounds of the corresponding effective
theory at $r\sim \lambda$, 
the space dimension of which is not necessarily that of the
fundamental theory.

Figure~\ref{fig1}  shows that for $\delta > 2$
particle-physics bounds, in particular 
those coming from collider physics, are
various orders of magnitude stronger than direct searches for 
non-Newtonian gravity below the mm. 
In other words, if any deviation is ever found
in these experiments, it
will not be possible to explain it in terms of large extra-dimension models.

On the contrary, for $\delta=1$ in the range $\lambda \agt 1$ nm
Casimir and Cavendish-like experiments are the most sensitive
and rule out a large amount of parameter space, while particle physics
is relevant only at much shorter distances. Notice that the bounds
still allow a strong gravity coupling (of the order of $1/({\rm TeV})^3$)
up to few hundred $\mu$m as long as it
then decreases fast enough to match the long distance regime. This
possibility could be important in the framework of
sterile neutrino physics~\cite{bulk}.

The case of $\delta =2$ is special and is discussed below.

The comparison depicted in Fig.~\ref{fig1} seems to suggest
that the most effective way of testing models with large
extra dimensions in experiments of non-Newtonian gravity is not by
going to shorter and shorter distances in the nm regime
by means of Casimir experiments but rather to 
improve the sensitivity of Cavendish-like experiments just below the
mm regime or Casimir experiments at the $\mu$m. For work in
progress in this directions and the first results, see~\cite{price}.
\vskip1.5em
{\bf Non-Newtonian gravity and compactification-dependent bounds.}
Once a particular compactification scheme
is chosen, bounds coming from non-Newtonian 
gravity tests can become more stringent because of the
relationship between $R^*$ and $M_f$. As we shall see, this is the
case for $\delta \leq 2$.

The comparison must be performed independently for the two 
ranges $r^* < R^*$ (SD)  and $r^* \agt R^*$ (LD), where the  predicted
non-Newtonian behaviors are different.

In the LD regime, the bound is obtained by finding the intersection of
 the stright line corresponding to $\alpha_G = \alpha_{\delta}$ 
with the bound coming from Cavendish-like 
experiments. The value of $\lambda$ at the intersection is the upper bound 
on $R^*$, from which one obtains a lower bound (model dependent) on $M_f$.

Each model is one point in the $(\alpha,\lambda)$-plane.
 Changing the shape of
the compactification manifold implies only an $O(1)$ correction factor 
to these bounds~\cite{expo}. 

The simplest example is depicted in
Fig.~\ref{fig2}, where the model of Ref.~\cite{add} (ADD) is used and
a specific potential derived from compactification 
on a $\delta$-torus. In this case we have 
\be
M_f^{2+\delta}= \frac{M_{\rm P}^2}{(R^*)^{\delta}} \, . \label{radius}
\ee  
The resulting bounds  are 
 $M_f \agt 5.3 \times 10^{5}$ TeV for $\delta = 1$, 
$M_f \agt 3.7$ TeV and $M_f \agt 3.1$ GeV, respectively, 
in  $\delta = 2$ and 3. 
The bound for $\delta = 1$ is much stronger than 
those from particle physics and often quoted to exclude altogether $\delta =1$
as an observable case. This conclusion however depends
on a specific way of treating the extra dimensions.      

In the SD regime, things are different: 
the experiment and the potential~(\ref{pot_s}) must be compared
at the distance explored by the experiment.
This comparison 
requires computing the force in the actual set-up of the
apparatus, and yields a direct bound on $M_f$. We have checked that
these bounds,  
derived in the Casimir  experiment~\cite{casimir2},
are two orders of magnitude weaker than those found in the LD
regime. For these experiments to be competitive in testing large extra 
dimensions, a further improvement of many orders of magnitude in 
sensitivity would be necessary. Therefore, 
we use the results of the LD regime to redo our analysis
and obtain the exclusion regions, shown in Fig.~\ref{fig3}, in the
case of compactification-dependent bounds.

More complicated compactification schemes, in particular those in which
different extra dimensions are treated in different manners, can be
discussed along the same lines by means of the corresponding effective
theories.

Contrarily to the case of particle-physics bounds, we have now weaker
bounds for larger numbers of extra dimensions. Only the cases $\delta
=$ 1 and 2 are more stringent than in the previous case.

\vskip1.5em
{\bf The case of two large extra dimensions.}
The case of $\delta =2$ is particularly interesting because,
as shown in Fig.~\ref{fig3}, the
different approaches give comparable bounds.
In the model dependent case,
the exclusion region is given by the area above the curve
denoted ADD in Fig.~\ref{fig4}.

Particle physics provides very stringent model-independent bounds
in the whole range; the best of them is denoted by Tevatron in
Fig.~\ref{fig4}. In the same figure,  for reference,
we also included the bound coming from
the measurament of the anomalous magnetic moment of the
muon ($\mu$AMM). 

In the near future,
LHC is expected to reach a sensitivity similar or better than that of the
model-dependent analysis ADD (see the curve LHC in Fig.~\ref{fig4}).  
In confronting non-Newtonian gravity experiments,
LHC will be competitive over the whole range of distances
except in the mm region where
an improvement of the sensitivity of Cavendish-like
experiments could be an important source of information.

We must also bear in mind that the case $\delta =2$ is strongly
disfavoured by astrophysical constraints~\cite{astro}.

\vskip1.5em
\begin{acknowledgments}
We thanks L.~Sorbo for comments on the manuscript.
\end{acknowledgments}  

\end{document}